\documentclass[a4paper]{cas-dc}

\usepackage[backend=biber,style=numeric,sorting=none]{biblatex} 
\usepackage[version=4]{mhchem}
\usepackage{amsmath}
\usepackage{subcaption}
\usepackage{graphicx}
\graphicspath{{figs/}}

\begin{document}
\let\WriteBookmarks\relax
\def\floatpagepagefraction{1}
\def\textpagefraction{.001}

\shorttitle{Heated Injector Design}
\shortauthors{A. Verma, J. Mango, S. Fukaya, A. Jadbabaie, S. Ebadi, R.F. Garcia Ruiz, J. M. Doyle}

\title [mode = title]{Design and Performance of a Heated Gas Injector for Producing Cold Molecular Beams}

\author[1,2,3]{Avneesh Verma}
\author[1,2]{Jack Mango}
\author[1,2,4]{Shungo Fukaya}
\author[1,2,4]{Arian Jadbabaie}
\cormark[1]
\cortext[cor1]{Corresponding author.} \ead{arianjad@mit.edu}
\author[1,2,4]{Sepehr Ebadi}
\author[1,4]{Ronald F. Garcia Ruiz}
\author[1,2]{John M. Doyle}

\affiliation[1]{organization={Harvard/MIT Center for Ultracold Atoms},
               city={Cambridge},
               state={MA},
               statesep={},
               postcode={02138}, 
               country={United States of America}}

\affiliation[2]{organization={Department of Physics, Harvard University},
               city={Cambridge},
               state={MA},
               statesep={},
               postcode={02138}, 
               country={United States of America}}

\affiliation[3]{organization={Division of Engineering Science, University of Toronto},
               city={Toronto},
               state={ON},
               statesep={},
               postcode={M5S 2E4}, 
               country={Canada}}

\affiliation[4]{organization={Department of Physics, Massachusetts Institute of Technology},
               city={Cambridge},
               state={MA},
               statesep={},
               postcode={02139}, 
               country={United States of America}}

\begin{abstract}
We realize an injector device that supplies warm gas directly into a cryogenic environment. This injector has several advantageous features, including robustness, rigidity, simple installation, and excellent thermal isolation between a hot ($\sim$300~K) copper fill line and a cold ($<$3~K) cryogenic buffer gas cell. Less than 200~mW heat load on the cell is observed in realistic conditions of a molecular precision measurement experiment. A polyamide-imide (PAI) tube is the essential design feature. The fill line is epoxied to one end of the tube while the other end of the tube is connected to the cell via a slip-fit onto a brass nipple, realizing a complete vacuum-tight seal. PAI contracts on the brass nipple when cooled, forming a cryogenic leak-tight seal. The injector is easily (de-)mountable and rigid, with no significant displacement of the fill line relative to the cell observed during cooldown to 4 K. We characterize injector performance by flowing into the cell \ce{SF6} through the hot fill line and cold \ce{He} buffer gas through a separate cryogenic fill line while laser ablating a barium-containing target. This produces cold BaF free radicals, detected using absorption spectroscopy. This injector design will be employed to laser cool radium-containing molecules, such as RaF and RaOH, where leak-tight delivery of \ce{SF6} and \ce{H2O} reagents into a cryogenic buffer gas cell is required for scientific and safety reasons. These molecules are of particular interest for the study of symmetry-violating nuclear properties and searches for physics beyond the Standard Model.
\end{abstract}


\maketitle

\section{Introduction}
Experiments employing cold molecular beams generated by cryogenic buffer gas beam sources (CBGBs) can require delivery of warm reactant gases directly into a cell filled with helium or neon cryogenic buffer gas~\cite{Brightguidedmolecular,BufferGasBeam,Wright2023cryogenic}. The reactants, $A$ and $B$, are cooled by collisions with the buffer gas and then react to form the target molecule $C$, either through collisions alone~\cite{chemistryAlFCaF,sunChemistryCryogenicBuffer2026} or with the aid of light-induced chemistry~\cite{Enhancedmolecularyield}. $A$ is typically introduced into the cell by a fill line while $B$ is typically introduced via laser ablation of a solid target within the cell. Product molecules $C$ thermalize with the cryogenic buffer gas and then exit the cell through a small ($\sim$ 3--7~mm diameter) aperture, resulting in a cold ($\sim$4~K), slow, and bright molecular beam in a vacuum region. This technique has been applied to produce a variety of molecular species for precision searches of physics beyond the Standard Model \cite{acmecollaborationImprovedLimitElectric2018,whiteSlowMolecularBeams2024} and as a starting point for laser cooling and trapping of molecules at temperatures $T\ll$ mK \cite{tarbuttLaserCoolingMolecules2018,fitchLasercooledMolecules2021}.

The freezing point of $A$ is almost always much higher than the temperature of the CBGB cell, so the fill line feeding $A$ into the cell must be kept hot, all the way into the cryogenic buffer gas volume inside the cell \cite{Brightguidedmolecular,chemistryAlFCaF}. Key requirements include a leak-tight path into the cell interior and a low enough heat load to keep the buffer gas cold. In addition, to avoid heating the cell, the heated fill line needs to enter the cell without touching its cold interior surfaces, thus requiring precise mechanical alignment. The assembly that aims to accomplish all of this is referred to as the ``injector''. 

We report the realization of a rigid injector that provides a leak-tight supply of warm reagent gas to a CBGB cell. We demonstrate the injector by producing a cold beam of barium monofluoride, BaF ($C$), formed in-cell from \ce{SF6} ($A$) and barium atoms ablated from a \ce{BaTiO3} target ($B$), with helium as the buffer gas. Our injector maintains precise dimensional stability; we observe no significant ($<1$ mm) displacement of the injector relative to cell during the pumping and cool down of the apparatus. 
With the fill line at 300 K, we measure a heat load of $166 \pm 10$ mW on the cell, well within the thermal ``budget'' of a typical CBGB source. The injector slip fits onto the cell and thus can be easily installed and replaced, either in situ (with the cell already mounted in a cryostat) or prior to cell installation.

\section{Injector Design}
\begin{figure*}
	\centering
	\includegraphics[width=\textwidth]{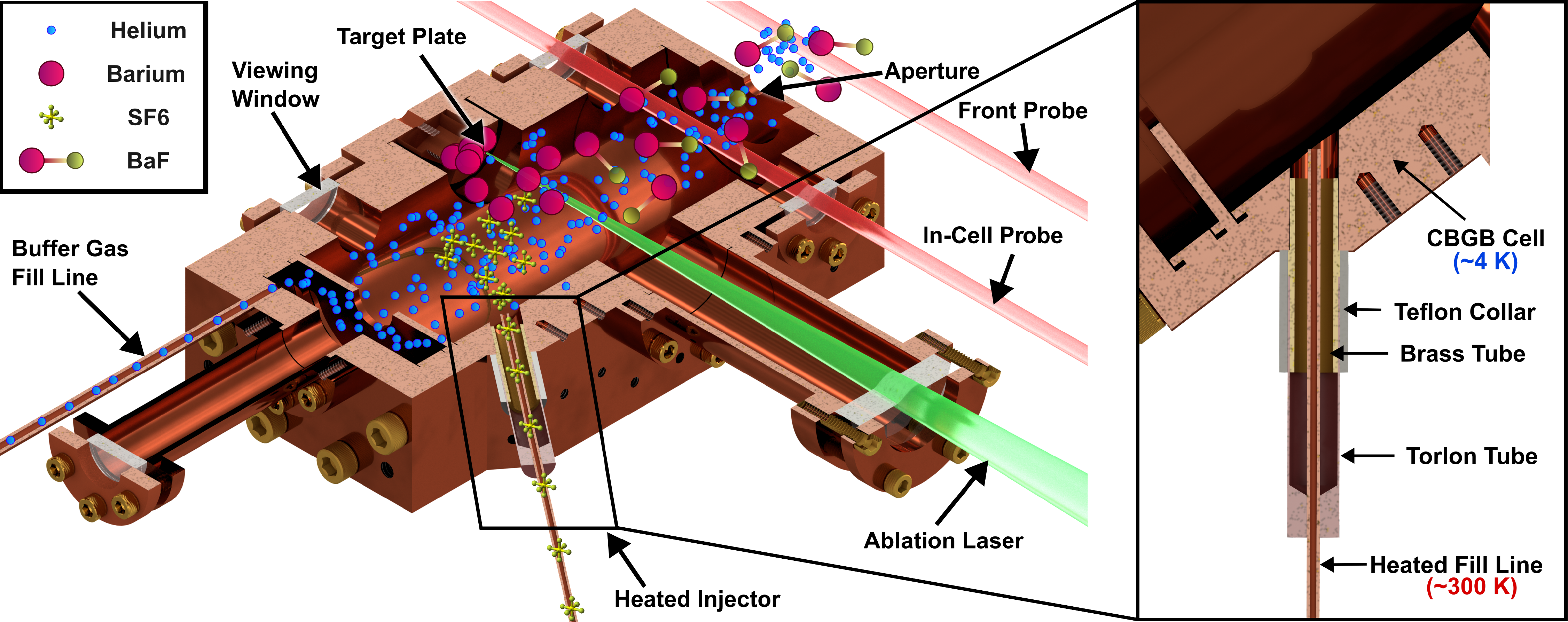}
	\caption{Main: Schematic cross-section of the Cryogenic Buffer Gas Cell (CBGB) setup. Helium buffer gas is introduced via a cryogenic fill line. A solid \ce{BaTiO3} target is laser ablated at 532 nm to produce Ba atoms, which react with \ce{SF6} gas introduced via the heated injector. This results in cold BaF molecules extracted out of the aperture into a beam by the buffer gas flow. Production and thermalization of BaF is monitored using absorption spectroscopy in-cell and in-front-of-cell. Inset: Close-up of the heated injector interface designed to isolate the ~300 K fill line from the ~4 K cryogenic cell body. The thin-walled PAI (Torlon) tube forms a leak-tight slip-fit over the brass tube, which is soldered to the copper cell. An optional PTFE (Teflon) collar adds additional structural rigidity and support.}
	\label{FIG:CellCrossSection}
\end{figure*}

A half-section view of the CBGB cell, including the injector, is shown in Figure \ref{FIG:CellCrossSection}. A copper (C12200) tube with 0.063" outer diameter and 0.014" wall thickness serves as the fill line. Although cryogenic apparatuses often use high-purity copper (C10100), the less pure C12200 alloy is satisfactory for a heated fill line since it will not be held at cryogenic temperatures, is more readily available for smaller diameter tubes, and is more rigid than annealed C10100 alloy. Four resistors\footnote{1 k$\Omega$ MP930 resistors} are mounted along the fill line and ohmically heat the line to $T_{FL} \sim$ 200--300~K. Typical heating currents are 15--30~mA, resulting in a nominal 250--900~mW heat load for each resistor. The end of the fill line fits through a hole on one end of a Polyamide-imide (PAI)\footnote{Torlon 4203} tube, and is sealed there with epoxy\footnote{Stycast 2850FT hardened with CAT 24LV}. The other end of the PAI tube slides onto a brass (C360) tube soldered\footnote{Sn/Pb 60/40 alloy solder} onto the CBGB cell\footnote{With a 0.001--0.002 inch radial gap between the surface of the brass tube and the hole in the cell to give space for the solder to flow}, providing a rigid leak-tight seal. Vacuum grease\footnote{Apiezon N grease} is applied to the surface of the brass tube before sliding on the PAI tube in order to enhance sealing. 

Tight tolerances and a smooth surface are found to be required to ensure a good slip-fit seal between the PAI and the brass tube on the CBGB cell. Brass was chosen since it is a relatively hard metal (compared to C10100 copper, which is the material of the CBGB cell), allowing easier machining of a smooth surface satisfying the tight tolerance constraints. The section of the PAI tube that slides onto the brass tube has an inner diameter of $d_T = 0.2300 \pm 0.0005$", while the brass tube has a matching outer diameter of $d_B = 0.2300 \pm 0.0005$". The PAI tube stretches slightly as it slips onto the brass tube. The fill line enters through the other end of the PAI tube, which has a $d_c = 0.0630 \pm 0.0005$" diameter hole with 0.001"--0.002" gap for epoxy to flow.

PAI was chosen for its rigidity and low thermal conductivity \cite{ThermalConductivityTorlon}, which minimizes the heat load on the cooled cell. In particular, PAI has higher stiffness and lower thermal conductivity than PTFE~\cite{NISTCryoMaterials}, which had been commonly used in heated injectors for CBGB cells in the past~\cite{BaoThesis, LoicThesis, HiroThesis}. Like PTFE, PAI has the crucial property of a much larger thermal contraction coefficient than metals, including the brass used here~\cite{Thermalexpansiontorlon, artcryogenicslowtemperature}, thus enhancing the plastic to metal seal when cooled. 

Brass has a larger thermal contraction coefficient than copper, which causes the brass tube to shrink slightly inside the hole in the cell to which it is soldered, resulting in a strain on the solder. This could cause the solder to crack at cryogenic temperatures \cite{Measurementmechanical4k}, but dunk testing the joint in liquid helium (discussed in Section \ref{sec:DunkTesting}) showed no observable leak for Sn/Pb~60/40 solder, as is typical experience for cryogenic seals of this size.

The design of our injector includes an optional PTFE (Teflon) collar that can be slid around the PAI-brass joint.  PTFE has a larger thermal contraction coefficient than both PAI and brass. Thus, a PTFE collar would tighten around the PAI-brass sandwich and the brass tube, adding additional leak-tight robustness to the injector at low temperatures. This potentially useful feature---especially for use in closed sealed cells containing hazardous materials---was not implemented in these tests.

\section{Helium Dunk Testing} \label{sec:DunkTesting}
\begin{figure*}
	\centering
	\includegraphics[width=\textwidth]{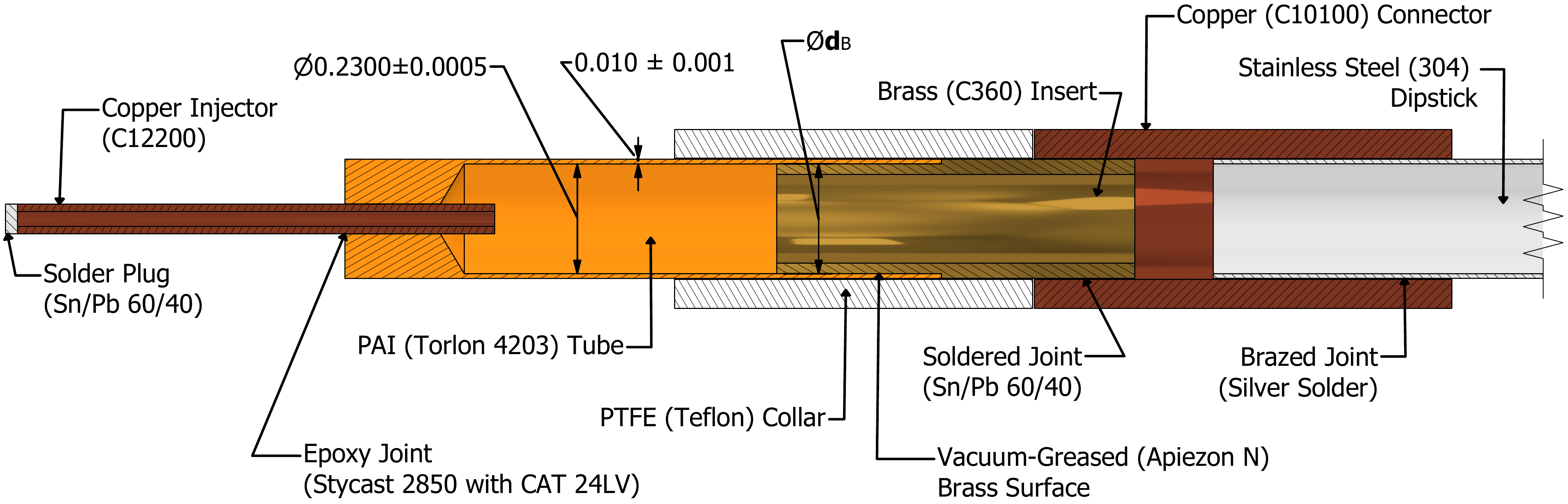}
	\caption{Cross-section view of the dunk test apparatus, detailing the material interfaces and dimensions of the heated injector assembly.}
	\label{FIG:DunkTestAssembly}
\end{figure*}

The reliability and sealing performance of brass-PAI interfaces and the copper-PAI epoxy joints are tested by leak-checking an injector assembly while repeatedly dunking it into liquid helium 7 times. A cross-section view of the ``dunk test'' of the injector is shown in Figure \ref{FIG:DunkTestAssembly}. This dunk test injector version matches the injector design described earlier, but with the copper fill line soldered closed to allow for easy leak checking of the seals. A copper connector, which plays the role of the CBGB cell, is brazed onto a 150~cm long thin-walled stainless steel tube (6.35~mm diameter), and then a brass tube is soldered onto the copper connector. The other end of the stainless steel tube is connectorized with a VCR fitting, which is connected to a helium leak detector\footnote{Pfieffer ASM 390} during testing. The injector end of this test assembly is dipped directly into a liquid helium storage dewar while helium leak detecting.

We note that in early designs of the test assembly, a copper tube brazed onto the stainless steel dipstick was used as the sealing surface for the PAI tube instead of brass, which is soldered to a copper surface. However, the sealing performance of this design was found to be unreliable, possibly due to residual pitting from brazing leaving the copper surface rough. Thus, a design involving the soldered brass tube was pursued. Brass has higher hardness than C10100 copper and we observe no pitting when soldered using Sn/Pb~60/40 alloy.

The following procedure was followed when conducting these ``dunk tests'', with the helium leak rate being logged at 1~Hz:

\begin{enumerate}
    \item Apply vacuum grease to brass tube.\footnote{One of the seven dunk tests was performed with neither vacuum grease applied nor the PTFE tube installed. The lack of vacuum grease had no discernible influence on the leak rate when dunked in liquid helium, verifying the robustness of the seal.}
    \item Slip PAI tube onto brass tube.
    \item Slide Teflon collar onto PAI and brass joint\footnote{Done in one dunk test, not employed in beam tests described in Section~\ref{sec:BaFTests}}
    \item Pump assembly to vacuum ($\sim$9~mTorr).
    \item While at room temperature, spray helium gas onto PAI and brass joint to check for a leak. 
    \item Dunk assembly into liquid helium dewar while leak testing.
    \item Once the test assembly is thermalized with the liquid helium, extract it from the dewar and spray with helium again.
    \item Vent assembly to bring inside back to atmospheric pressure. Remove PAI piece from brass tube.
    \item Repeat, starting at step 2.
\end{enumerate}

The procedure was repeated seven times using a brass piece of $d_B=0.2300 \pm 0.0005$". The diameter $d_B$ was tuned through multiple machining runs to give a ``snug'' fit at room temperature, providing good orientational stability of the PAI tube relative to the brass tube. In all dunk tests, no direct leaks above $1 \times 10^{-7}$~mbar l/s of helium were observed, either at room temperature or after thermalization in liquid helium. However, occasional small spikes in the helium leak rate were observed whenever the assembly was bumped while dunking or extracting it from the dewar. Small increases in the leak rate ($\sim 3 \times 10^{-8}$  mbar l/s) were observed over long times at room temperature, though not at low temperatures, characteristic of helium diffusion through the wall of the PAI tube \cite{TemperatureBehaviourPermeation}.

\section{Demonstration of Cold Molecular Beam Production}\label{sec:BaFTests}
The cryogenic buffer gas cell shown in Figure \ref{FIG:CellCrossSection} implements this heated injector design, and will be used to supply reagent gas to produce cold beams of laser-coolable molecular free radicals. For initial commissioning, the cell was loaded with a barium titanate (\ce{BaTiO3}) ablation target to demonstrate successful production of \ce{BaF} using a heated flow of \ce{SF6}. In these tests, the injector was initially run at 232~K. This caused the cell temperature to rise from its baseline at 2.18~K to 2.40~K, which corresponds to a $98.5 \pm 0.8$~mW heat load on the cell, calibrated with independent measurements. Upon flowing helium buffer gas at 3.92~sccm, the equilibrium cell temperature increased to $\sim$2.51~K, corresponding to an additional steady state heat load of $49.2 \pm 0.4$~mW from buffer gas collisions with the hot fill line. These collisions also caused the injector's steady state temperature with these heater parameters to drop from 232~K to 226~K.

\begin{figure*}
    \centering

    \begin{subfigure}[t]{0.48\textwidth}
        \centering
        \includegraphics[width=\textwidth]{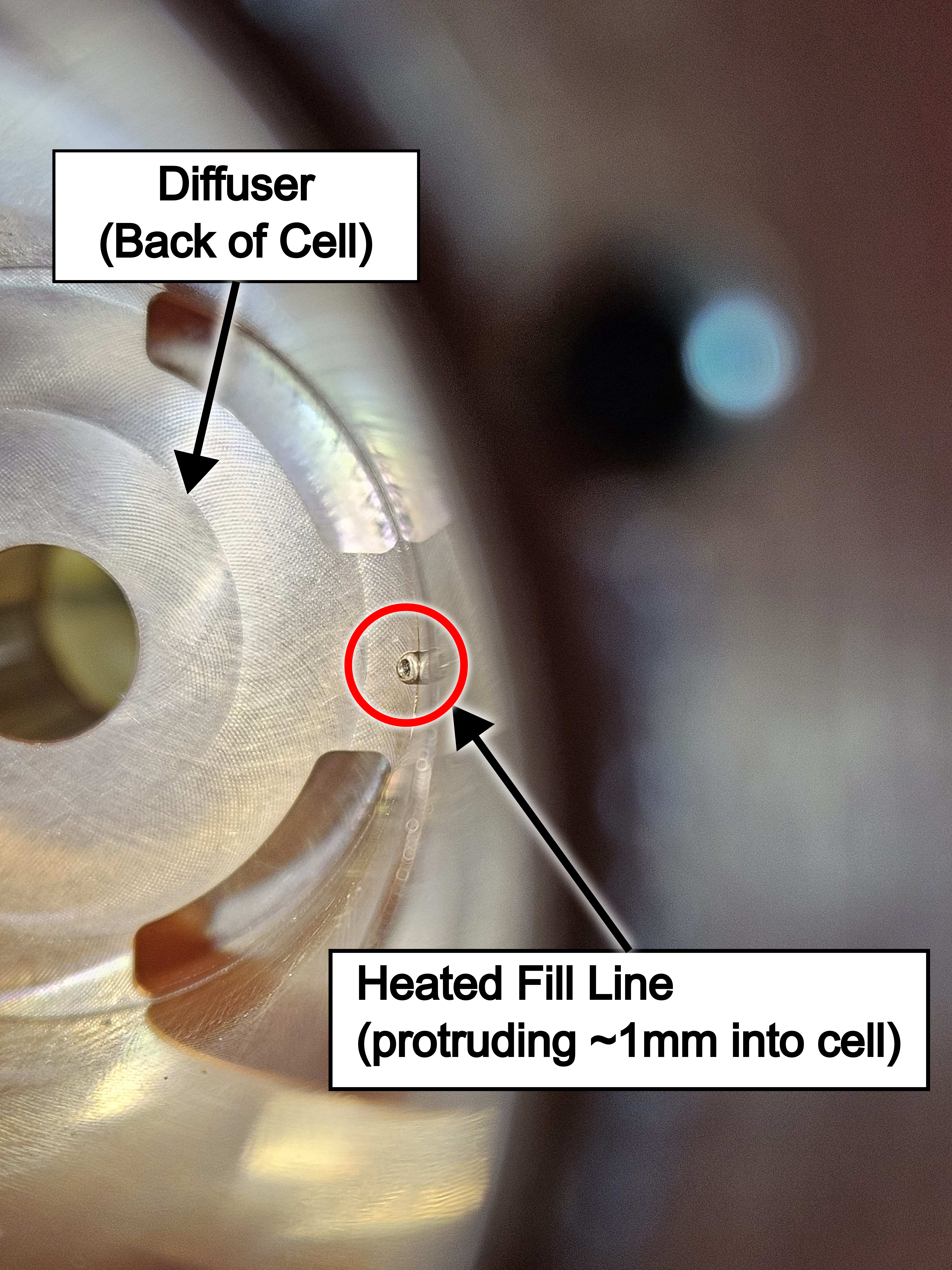}
        \caption{View down the bore of the cell (front of cell disassembled), showing the heated injector protruding into the cell bore. The injector protrudes as into the cell as little as possible to minimize heating of the cell, yet just enough to avoid spraying ligant gas onto the walls of its hole.}
        \label{fig:frontview}
    \end{subfigure}
    \hfill
    \begin{subfigure}[t]{0.48\textwidth}
        \centering
        \includegraphics[width=\textwidth]{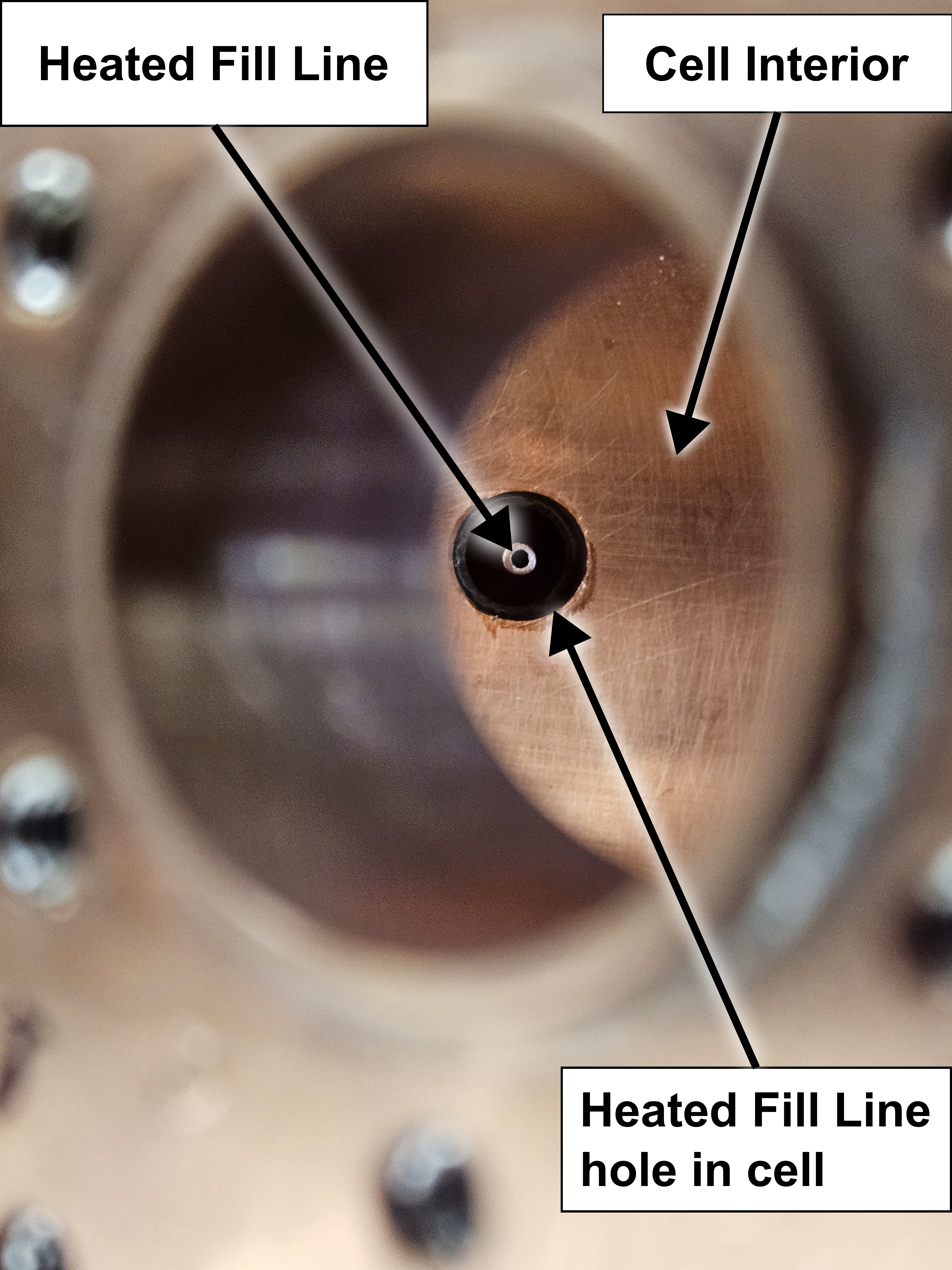}
        \caption{A view of the heated fill line from the hole where the surface holding the targets would sit. The heated fill line is centered in the hole in the cell that it feeds through.}
        \label{fig:targetview}
    \end{subfigure}

    \caption{Images of the inside of the CBGB cell, with the heated injector installed.}
    \label{fig:heated-injector-design}
\end{figure*}

These heat loads are well within the capacity of cryocoolers used for CBGB cells. For reference, the heat load due to thermal conductivity through the PAI (a.k.a. Torlon) is expected to be $\sim$10~mW, and the heat load from blackbody radiation is estimated to be up to $\sim$80~mW from the portion of the heated copper tube surrounded by the cell wall. The rest of the heat load is hypothesized to be due to blackbody radiation from the portion of the heated fill line leading up to the Torlon tube (which was wrapped with 4 layers of aluminized mylar to reduce radiative heat loads). 

A window placed opposite to the heated injector allowed the injector to be monitored with a camera. Crucially, any thermal short between the  copper injector line and cell will rapidly cool the injector and freeze the reagent gas before it reaches the cell. Thermal shorts can occur either by injector deflection causing contact with the cell walls, or by the formation of reagent ice-bridges between the cell wall and injector exit. Initial observations from  the camera did not show either behavior; however, after a few ablation shots, the window was covered with dust from ablation, making it challenging to check for ice bridges. In the future, a copper spacer can be used to separate the window from the main bore of the cell, reducing the amount of dust that obscures the window.

The injector protrudes $\sim$1~mm into the bore of the cell, as seen in Figure \ref{fig:frontview}, which shows the view from the front of the cell looking down the bore. The injector must protrude into the cell just enough to ensure the sprayed reagent clears the cell wall. Still, protrusion into the cell is minimized, since increased length of heated copper inside the CBGB cell would result in additional heat load from buffer gas collisions. Figure \ref{fig:targetview} shows the injector centered in its hole, viewed from the target location. 

\begin{figure*}
    \centering
    \begin{subfigure}[t]{0.47\textwidth}
        \centering
        \includegraphics[width=\textwidth]{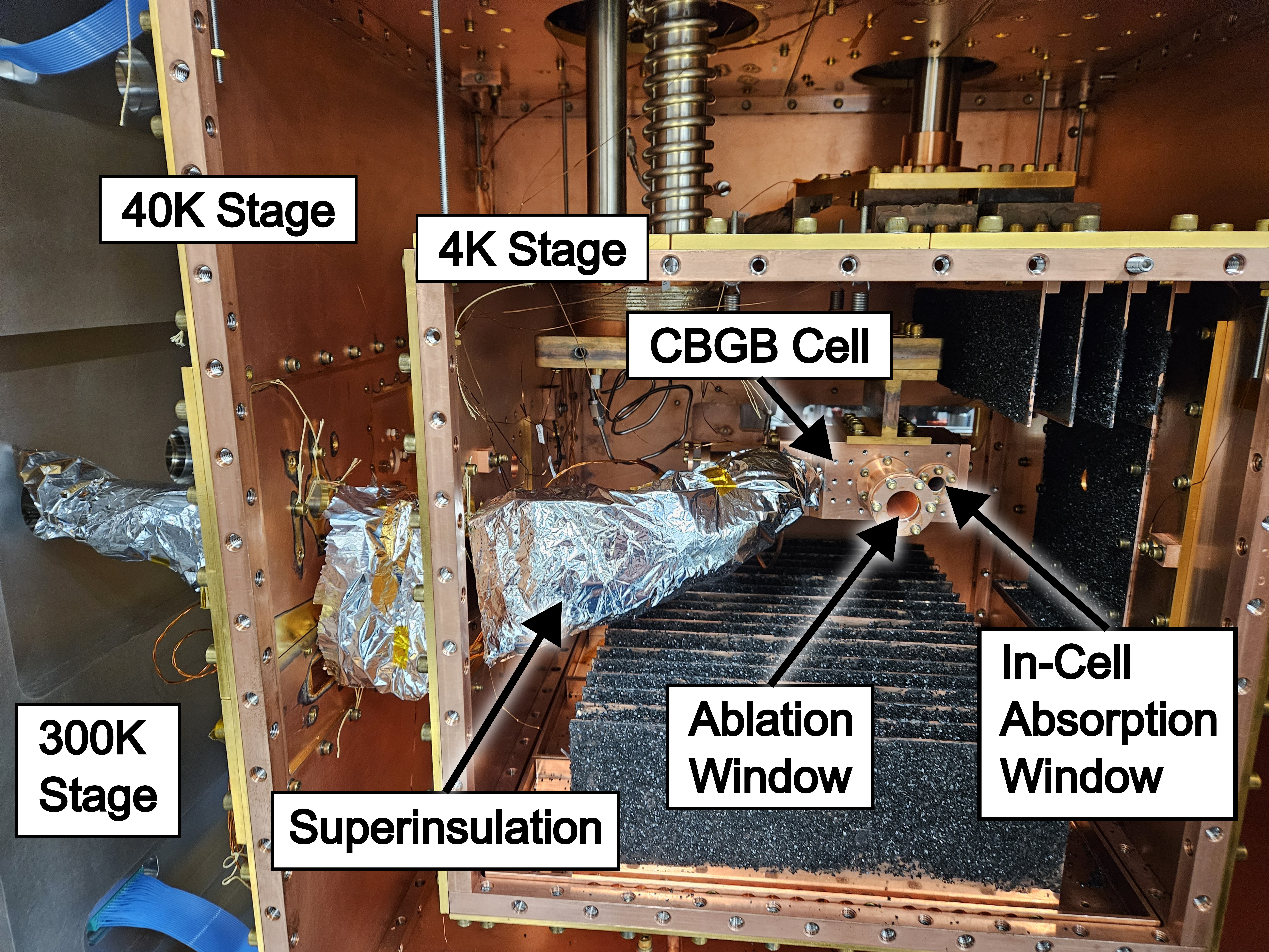}
        \caption{Heated Fill Line installed in CBGB source, wrapped with Superinsulation (aluminized mylar).}
        \label{fig:superinsulation}
    \end{subfigure}
    \hfill
    \begin{subfigure}[t]{0.47\textwidth} 
        \centering
        \includegraphics[width=\textwidth]{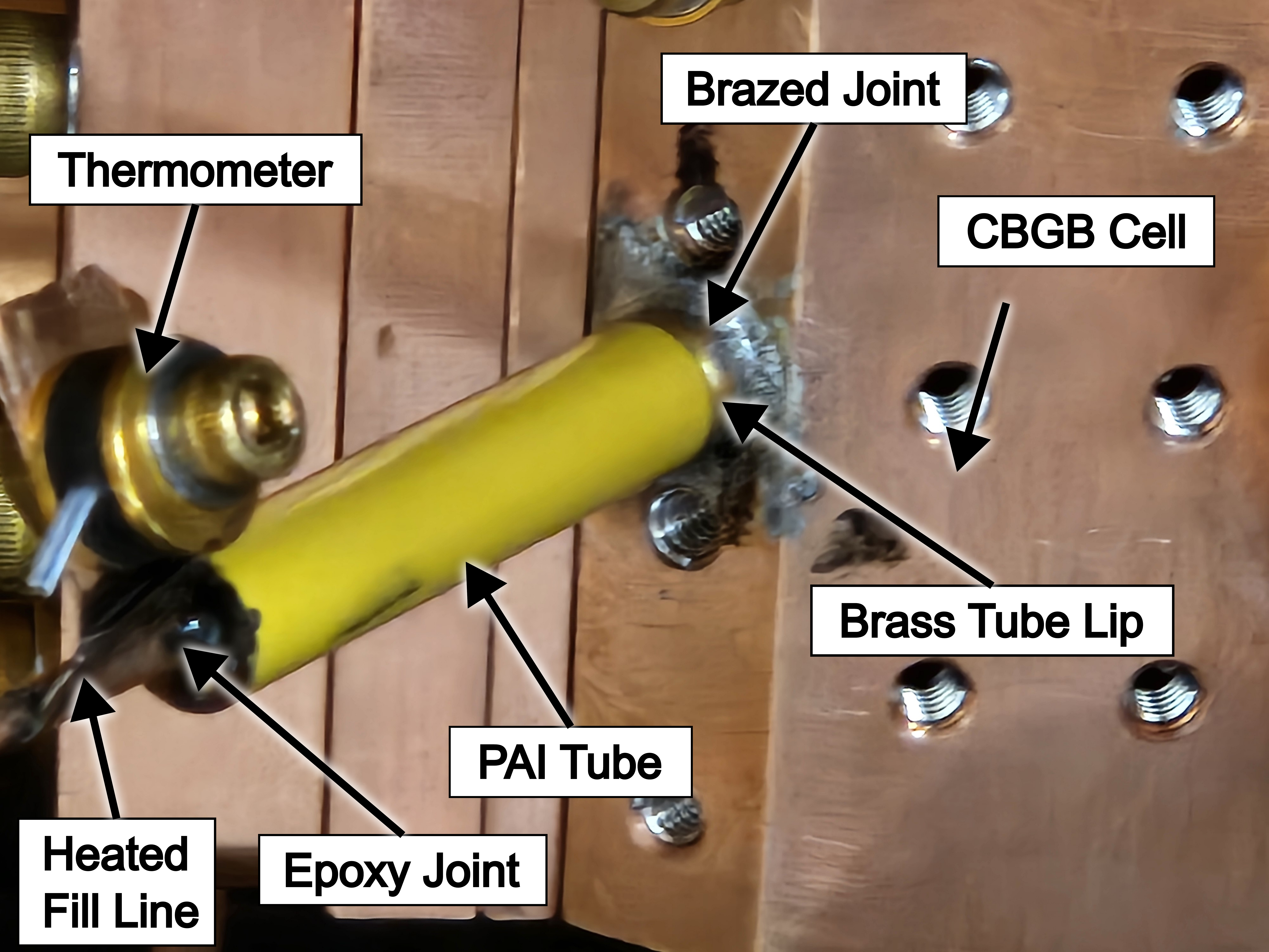}
        \caption{PAI Tube installed onto the CBGB Cell}
        \label{fig:InstalledPAI}
    \end{subfigure}
    \vfill
    \begin{subfigure}[t]{0.95\textwidth}
        \centering
        \includegraphics[width=0.75\textwidth]{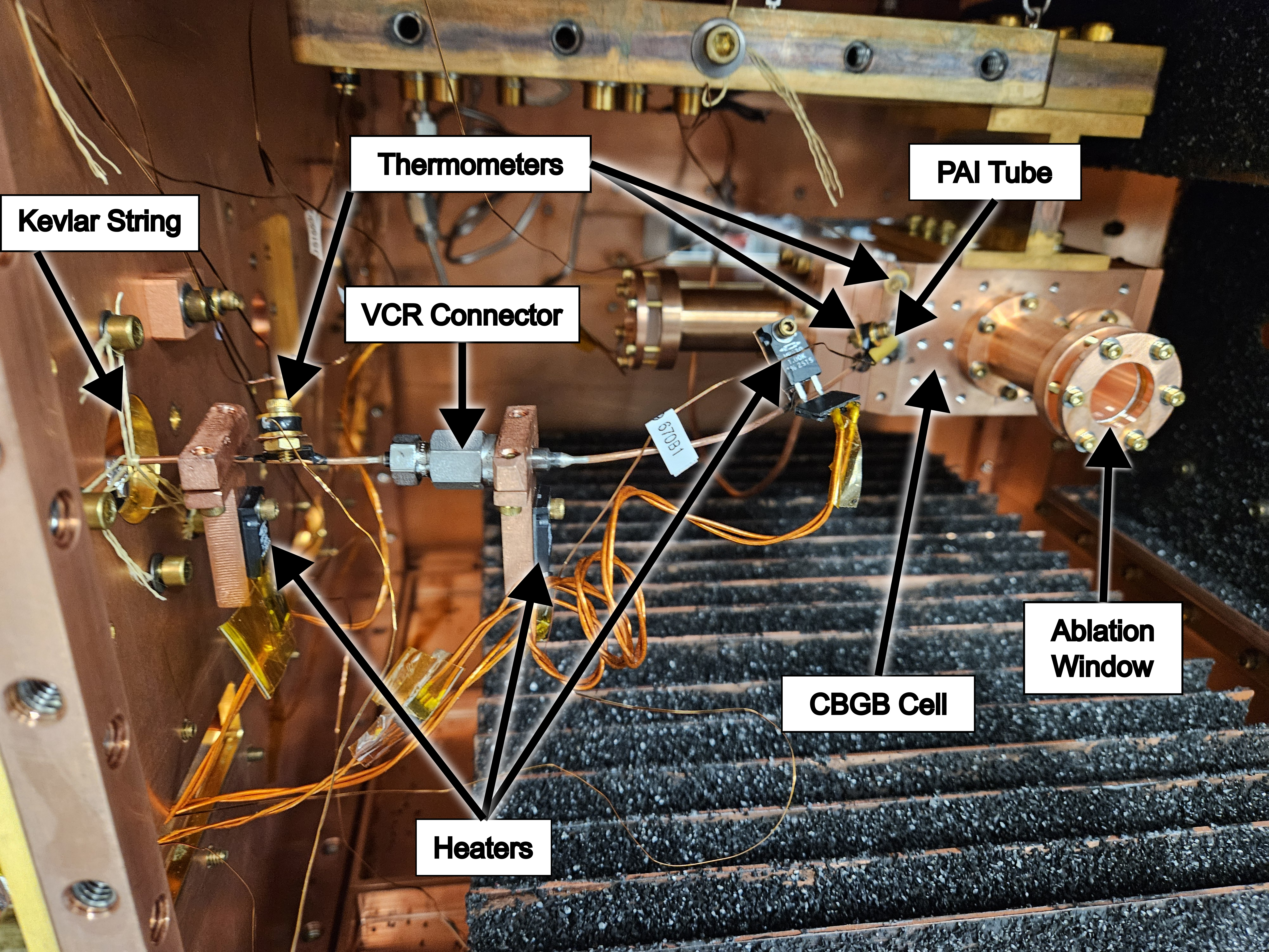}
        \caption{Heated Fill Line assembly installed in CBGB source, prior to wrapping with superinsulation.}
        \label{fig:HFLinBeambox}
    \end{subfigure}
    \caption{Heated Fill Line installed in CBGB Source. Visible at the bottom of the source are charcoal sorbs used to cryopump the helium buffer gas. }
    \label{fig:HFL}
\end{figure*}

The full heated fill line assembly in the cryogenic vacuum chamber is shown in Figure \ref{fig:HFL}. In particular, Figure \ref{fig:HFLinBeambox} shows the heated fill line, along with thermometers\footnote{Lakeshore DT670 Silicon Diode} and heaters mounted on copper clamps or epoxied to copper foil. A VCR joint modularly connects the injector that mounts onto the cell to the rest of the heated fill line (which mates with the feedthrough on the vacuum chamber wall). Kevlar string tied to brass screws on the 40~K and 4~K radiation shields support the heated fill line as it passes through holes in the shields. Figure \ref{fig:InstalledPAI} shows a close-up of the PAI tube of the injector slip fit onto the brass tube soldered to the CBGB cell, and Figure \ref{fig:superinsulation} shows the whole assembly after it is wrapped in aluminized mylar (to reduce blackbody radiation from the heated fill line).

To produce \ce{BaF}, the \ce{BaTiO3} target was ablated with $15$~mJ at 532 nm using a pulsed Nd:YAG  expanded to 15~mm diameter and focused with a 500 mm lens. Simultaneously, we flow helium buffer gas and \ce{SF6} reagents into the cell. For these tests, we used 0.18~sccm of \ce{SF6} flow via the heated fill line, and 3.92~sccm of helium flow via a separate fill line thermally anchored to the cell cold stage. As stated earlier, this buffer gas flow resulted in an additional steady-state thermal load of $\sim$50~mW on the cell.

The production and thermalization of \ce{BaF} was verified by absorption spectroscopy both inside and a few mm outside the cell, confirming the injector successfully supplied \ce{SF6} to react with the ablated barium. The absorption probes are derived from a single CW laser resonant with the $Q_1(0)$ line ($N''=0^+,J''=1/2\rightarrow J'=1/2^-$) of the $X^2\Sigma^+ (v''=0) \rightarrow A^2\Pi_{1/2}(v'=0)$ transition in $^{138}$BaF~\cite{PrecisionspectroscopyA2P-X2S, MolecularbeamopticalStark}. The laser frequency is monitored with a wavemeter\footnote{HighFinesse WS7-30}, locked to $\sim$1~MHz precision when at a fixed frequency, and referenced to a Rb reference laser with $\pm$500~kHz absolute accuracy\footnote{HighFinesse SLR 780}. When scanning, the laser lock has an error of $<5$~MHz. 

Absorption signals were converted to optical depth (OD), which is proportional to the density of molecules interacting with the probe laser beam. For a weak probe beam with a nominal intensity of $I_0$ transmitted through the cell, the presence of an absorbing sample will reduce the transmitted light to $I$, with the OD given by the Beer-Lambert law:
\begin{equation}\label{eq:OD}
    \text{OD} = \ln\left(\frac{I_0}{I}\right)=n\sigma l
\end{equation}
where $n$ is the number density of absorbers, $\sigma$ is the cross-section for an absorber interacting with the probe, and $l$ is the probe path length through the sample. 

To convert from OD to density $n$, we must consider the effect of Doppler broadening on the absorption cross-section, detailed in Refs.~\cite{Budker2008,LaserSpectroscopy1}. For BaF at $\sim$4~K, the Doppler-broadened full-width at half maximum (FWHM) is $\Gamma_D/2\pi \approx 40$~MHz, while the probe laser only interacts with molecules within a natural linewidth $\gamma/2\pi=3.46$~MHz~\cite{MolecularbeamopticalStark} of the transition. The resonant cross section is reduced by $\gamma/\Gamma_\text{tot}$, and for a Doppler-dominated Voigt lineshape we have $\Gamma_\text{tot}\approx \Gamma_D+\gamma/2$. We therefore obtain an estimate of $\sigma_0 \approx 6.1\times10^{-15}$~m$^2$ for the $Q_1(0)$ line. With the additional assumption the molecules are equally distributed along the $l=1.5$ probe laser path across the cell, we estimate the conversion factor of OD to density in-cell as $n_\text{cell}/\text{OD} \approx4.3 \times10^9~ \text{cm}^{-3}$. 


\begin{figure}
    \centering
    \includegraphics[width=\columnwidth]{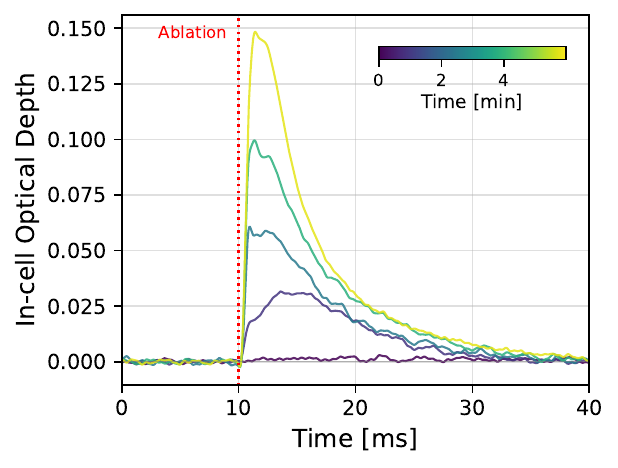}
    \caption{Optical depth for shots taken while the \ce{SF6} flow rate builds up from 0 to 0.18~sccm. Absorption signal is observed to increase as the \ce{SF6} density builds up in the CBGB cell.}
    \label{fig:SF6_switched}
\end{figure}

In Figure \ref{fig:SF6_switched}, we show the evolution of absorption signals as the \ce{SF6} flow is switched from off to on, while monitoring BaF production from $\sim$1~Hz ablation of the target. The first few shots show no absorption with the probe beam at the \ce{BaF} resonant frequency. Upon opening a valve to let in \ce{SF6} flow, the peak OD increases to a maximum value of $\sim$$0.15$ (corresponding to a \ce{BaF} density of $n_\text{cell}\approx6.5\times10^8$~cm$^{-3}$ in the probed level) as the reagent flow and in-cell density approaches steady state. This confirms the heated injector is successfully supplying \ce{SF6} gas at a sufficiently high temperature to reach the reaction region inside the CBGB cell without freezing. 

\begin{figure}
    \centering
    \includegraphics[width=\columnwidth]{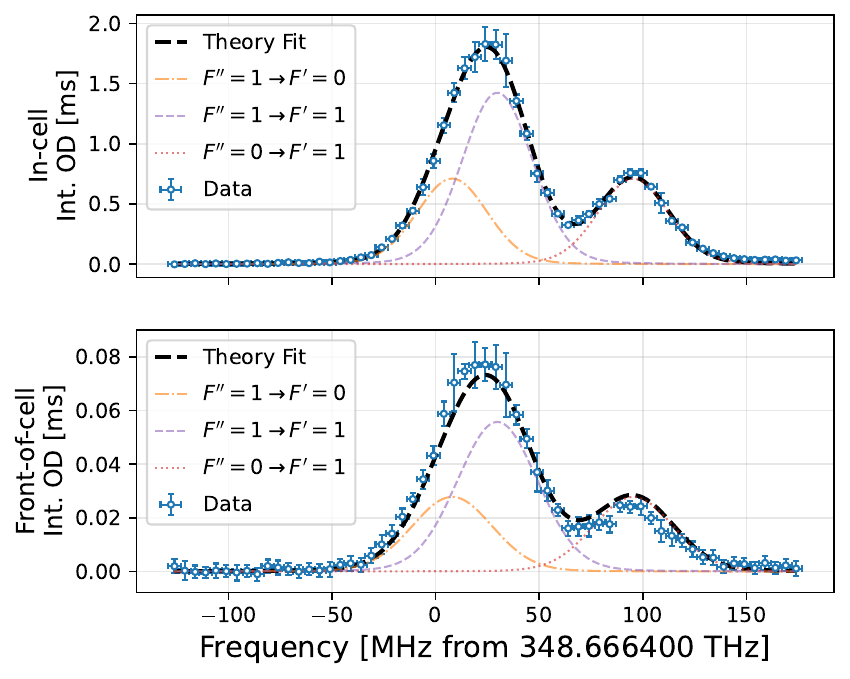}
    \caption{Integrated Optical Density over frequency scanned around the $^{138}$BaF $X^2\Sigma^+ (0) \rightarrow A^2\Pi_{1/2}(0)$ $Q_1(0)$ line. Two hyperfine sublevels from Fluorine nucleus are observed. The front of cell signals show lower optical depth since the density of molecules in front of the cell is limited by the aperture and the divergence of the beam from the region addressable by the probe beam diameter. See text for details. }
    \label{fig:spectrum}
\end{figure}

To further verify the cooling performance of the cryogenic buffer gas cell operating with the heated injector, we scan the frequency of the absorption laser over 300 MHz centered on the $Q_1(0)$ transition in $^{138}$\ce{BaF} \cite{MolecularbeamopticalStark}. For each frequency, the measured OD was integrated over the duration of the absorption signal, $25$~ms in-cell and $15$~ms in-front-of-cell. The resulting integrated OD is proportional to the total number of \ce{BaF} molecules in the ground states addressed by the probe laser, shown for both in-cell and in-front-of-cell signals in Figure \ref{fig:spectrum}. 

The observed spectra are consistent with previous high-resolution spectroscopy of $^{138}$BaF~\cite{PrecisionspectroscopyA2P-X2S, MolecularbeamopticalStark}. The observed peak splitting is predominantly due to the hyperfine interaction of the valence Ba-centered electron and the $^{19}$F nuclear spin in the ground electronic state. The excited hyperfine interaction (splitting $22$~MHz)~\cite{BenchmarkingFockspacecoupledcluster} results in three lines total in high-resolution beam spectra~\cite{PrecisionspectroscopyA2P-X2S}, but for in-cell spectra two of the lines are unresolved in the lower frequency peak. Simulations of line positions and strengths using effective Hamiltonian parameters obtained from previous spectroscopy~\cite{HighResSpecBaF} are in good agreement with the data (see Appendix \ref{sec:BaFsims} for details). 

The frequency lineshape of this spectra can provide information on the performance of the cryogenic buffer gas cooling in the cell. In Figure~\ref{fig:spectrum}, we fit the data to the combination of three Voigt distributions fixed at each predicted hyperfine line location and scaled by the predicted line strength. The Lorentzian width is fixed at the natural linewidth, while the Gaussian width $\Gamma_D$ is taken to be the same for all hyperfine lines, and is a free fit parameter. 

The in-cell fits converge to a Doppler width of $\Gamma_D/2\pi = 40\pm0.5~$MHz, corresponding to a $T=4\pm0.1$~K translational temperature averaged over the temporal window of $1-25$~ms. The observed peak widths of the data depend on the choice of integration window, as the hot molecules produced in ablation are thermalized to the cell temperature~\cite{BufferGasBeam}. By temporally binning the data and fits, we observe thermalization of the Doppler width during the molecular pulse duration in two timescales. The first timescale of $\sim\!150$~\textmu s corresponds to initial cooling of the ablation plasma from $T\gg4$~K to $T\sim4$~K, in agreement with previous studies~\cite{skoffDiffusionThermalizationOptical2011}. The second timescale of $\sim\!10$~ms corresponds to cell dynamics, such as diffusion and buffer gas flow, depleting hot molecules over time~\cite{BufferGasBeam}. After $\sim\!10$~ms, fit widths indicate $<3$~K temperature, consistent with the buffer gas cell temperature during ablation ($\sim\!2.6$~K).

The front-of-cell widths converge to $\Gamma_D/2\pi=46\pm2$~MHz, though this should not be interpreted as a temperature. The front-of-cell linewidths are further broadened by collisional boosting from buffer gas flow out of the cell aperture~\cite{Shapednozzlescryogenic}. Instead, by converting $\Gamma_D$ to velocity, we obtain a FWHM of $\Delta v_\perp\approx40$~m/s in the transverse direction to the beam. This is close to the $35$~m/s transverse velocity lower bound from an effusive source at $4$~K, with the additional velocity spread attributable to less than one buffer gas collision on average per molecule (i.e. $Re\lesssim1$)~\cite{BufferGasBeam}. Note that collisions and therefore boosting are expected to continue beyond the few mm distance of the probe beam, out to the $\sim$cm scale.

The total linewidth $\Gamma_\text{tot}\approx \Gamma_D +\gamma/2$ can be used to estimate the density of the molecules in front of the cell. The aperture of the cell is $5$~mm in diameter, and we take $l\approx7$~mm to account for transverse divergence in front of the cell~\cite{BufferGasBeam}. We therefore obtain $\sigma_0\approx 5\times10^{-15}\,\text{m}^2$ and a conversion of $n_\text{front}/\text{OD}\approx 3 \times 10^{10}\,\text{cm}^{-3}$. Given the observed temporal FWHM of $\sim$5~ms for the front-of-cell signal, we estimate the peak density from Figure~\ref{fig:spectrum} as $n_\text{front} \approx 5\times10^8\,\text{cm}^{-3}$ molecules in the $F^{\prime\prime}=1$ ground state.

\section{Conclusions}
We have demonstrated a heated gas injector that delivers warm reagent gas into a 4~K CBGB cell with a heat load of $166 \pm 10$~mW (no buffer gas flow) and $\lesssim 200$~mW under typical operating conditions. We have shown the injector remains leak-tight to better than $10^{-7}$~mbar$\cdot$L/s through repeated thermal cycling. The injector is rigid, easily mountable and demountable, and compatible with sealed cells at cryogenic temperatures. We demonstrated the performance of the injector by flowing \ce{SF6} gas at 230~K into the cell during laser ablation, successfully producing a cryogenic buffer gas beam of \ce{BaF} at 4~K. 

This heated injector design will be used in multiple experiments that benefit from leak-tight construction. In one experiment, a cold beam of \ce{CaNH2} free radicals will be produced by ablating \ce{Ca} and flowing \ce{NH3} reagent gas, where the leak-tight design ensures reactive ammonia does not contaminate the vacuum chamber and charcoal sorbs that pump helium emanating from the CBGB. The design will also be used in an experiment producing cryogenic buffer gas beams of \ce{^{226}RaF} and \ce{^{226}RaOH} free radicals for laser cooling toward searches for new physics~\cite{RadioactiveMoleculesLaboratories,CaltechRaX}. Similar to the demonstrated case here, the heated injector will supply \ce{SF6} or \ce{H2O} during ablation of a radium-containing target~\cite{CaltechRaX} to generate molecules of interest. The leak-tight injector design help mitigate the spread of radon gas (produced by decays of the $^{226}$Ra target loaded in the cell), and the removable PAI tube facilitates rapid prototyping and repeatability when managing radioactivity in the cell. 

\section*{Acknowledgments}
We thank Samuel Munoz Arias, Matteo Fulghieri, Karina Khusainova, and Soumitra Ganguly for assistance in fabricating the CBGB experiment, and Nicholas R. Hutzler for insightful discussions and feedback on the manuscript. This work was carried out as part of the RaX collaboration between Caltech, Harvard, MIT, and FRIB, whose members we thank for helpful discussions and advice in construction of the apparatus. We acknowledge  support from the U.S. Department of Energy, Office of Science, Office of Nuclear Physics under grants DE-SC0026217, and the Harvard–MIT Center for Ultracold Atoms (CUA) from the National Science Foundation. A.J. acknowledges support from the National Science Foundation (PHY-2402254).

\printcredits

\printbibliography
\appendix

\section{Diffusion Constant Estimation}

\begin{figure*}
	\centering
	\includegraphics[width=0.7\textwidth]{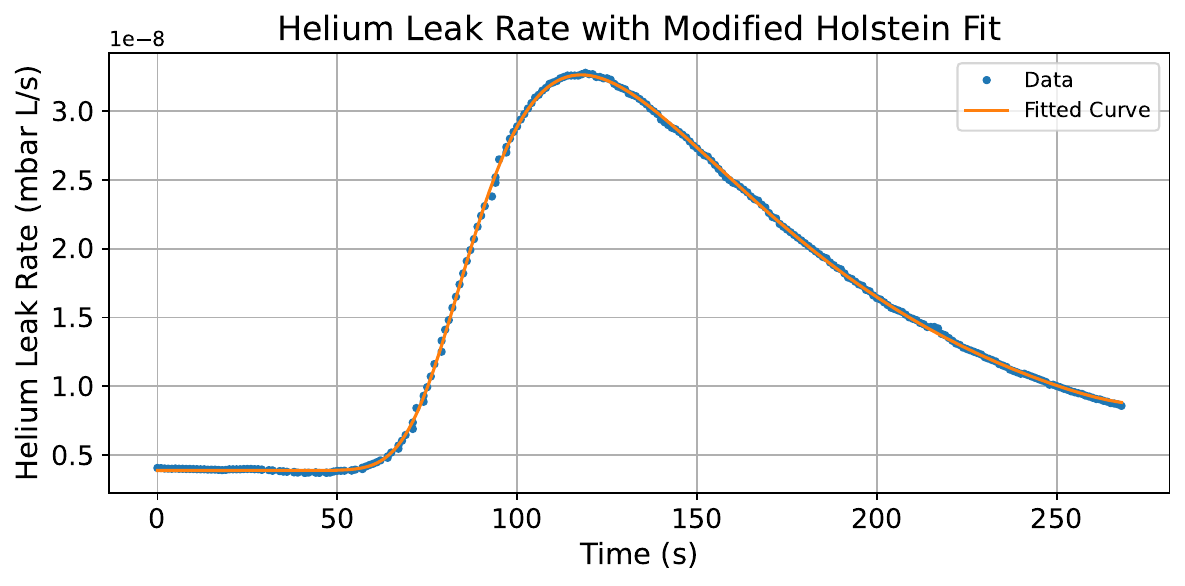}
	\caption{An example of leak rate data fit to the modified Holstein approximation given by Equation~\ref{EQ:fit_piecewise} convolved with a Gaussian.}
	\label{fig:DunkTestFit}
\end{figure*}

The helium diffusion constant $D$ through the PAI mount can be estimated using our leak rate data. When a helium concentration gradient is applied between two sides of a polymer, the solution to the diffusion equation can be modeled by Holstein's approximation \cite{TemperatureBehaviourPermeation, DiffusionCoefficientSolubility}:
\begin{equation} \label{EQ:Holstein}
    Q(t) = 2Q_\infty \left(\frac{\tau}{t}\right)^{\frac{1}{2}} \exp{\left(- \frac{\pi \tau}{4t}\right)}
\end{equation}
where $Q$ is the measured leak rate at time $t$ and $\tau$ is the diffusion time constant, which is given by
\begin{equation} 
    \label{EQ:tau}
    \tau = \frac{d^2}{\pi D}
\end{equation}
where $d$ is the wall thickness of the sample (in this case, the PAI tube) and $D$, as mentioned earlier, is the diffusion constant of the gas through the polymer (in this case, helium gas through PAI). 

Holstein's approximation applies to a constant concentration of gas on one side of the sample, and a vacuum on the other, whereas we apply a transient concentration of helium for a short time ($\sim$30--90 seconds). The leak rate as a function of time follows Holstein's approximation modulated by an exponential decay. Thus, the leak rate is fit to a Holstein approximation, but multiplied by an exponential decay that begins at a time $t_1$, which is left as a fit parameter. This is then convolved with a gaussian to provide a smooth function (whose standard deviation is also left as a fit parameter, and depends on the duration of the helium spray). We fit the measured leak rate to the convolution of the following function $f(t)$ with a Gaussian:
\begin{equation}\label{EQ:fit_piecewise}
    f(t) = \begin{cases}
        Q_0, & t \leq t_0 \\
        Q(t-t_0) + Q_0, & t_0 < t \leq t_1 \\
        Q(t-t_0)\exp\left(\frac{t - t_1}{\tau'}\right) + Q_0, & t \geq t_1
    \end{cases}
\end{equation}
where $Q_0$ is a constant offset representing the baseline leak rate, $t_0$ is the time when the leak rate begins to rise in accordance with Holstein's approximation, $t_1$ is the beginning of the exponential decay as mentioned earlier, and $\tau'$ is a time constant associated with the exponential decay, which in general may be different from the diffusion time constant $\tau$ mentioned earlier. All of these are left as fit parameters and calculated for various tests. An example of the observed leak rate vs time and the fitted curve is shown in Figure \ref{fig:DunkTestFit}.

The diffusion constant of helium through the PAI at room temperature was found to be $D=(2.6 \pm 1.2) \times 10^{-10} \, \mathrm{m}^2/\mathrm{s}$, which is roughly consistent with the literature value of $\sim1.2 \times 10^{-10} \, \mathrm{m}^2/\mathrm{s}$~\cite{TemperatureBehaviourPermeation}.

\section{Estimated Thermal Loads} \label{sec:HandCalcs}
The heat loads due to conduction through the PAI tube and blackbody radiation of the portion of the heated injector within the cell can be estimated by hand. 

\begin{figure*}
    \centering
    \includegraphics[width=0.9\textwidth]{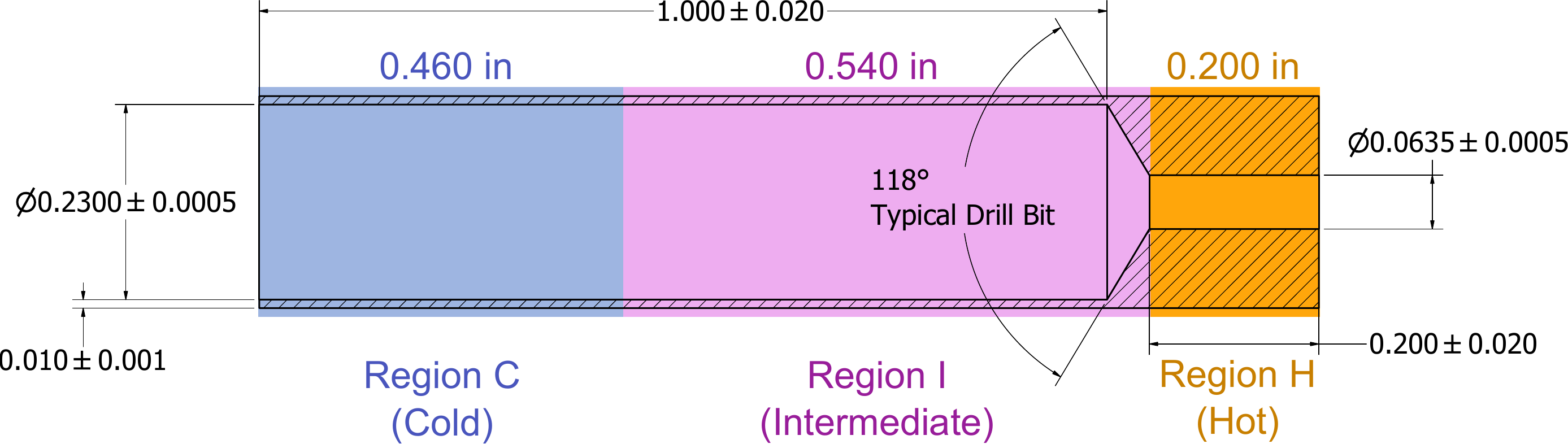}
    \caption{Engineering Drawing of the PAI tube, with regions labeled based on temperature. For thermal load estimates, Region H (the ``Hot'' region) is assumed to be at the temperature read by the injector thermometer just before the PAI tube (232~K without helium flow). Region C (the ``Cold'' region) is the portion of the PAI tube that slips onto the brass tube thermally anchored to the CBGB cell, and is thus assumed to be at the cell temperature (2.4~K without helium flow). Region I (the ``Intermediate'' region) is the portion of the PAI tube with a thermal gradient from the ``Hot'' side to the ``Cold'' side.}
    \label{fig:TorlonRegions}
\end{figure*}

\begin{figure}
    \centering
    \includegraphics[width=\columnwidth]{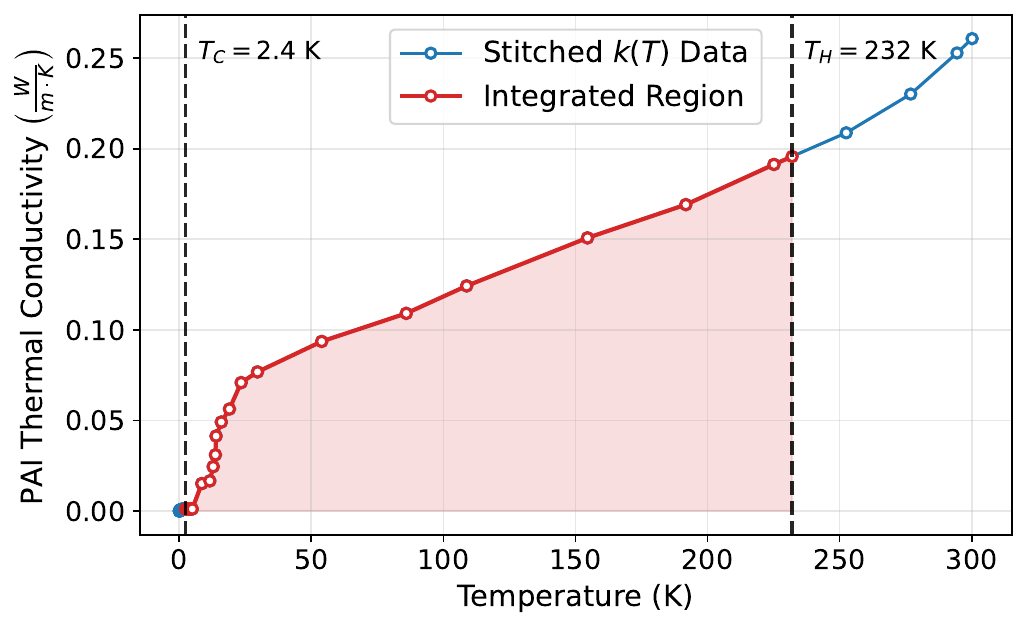}
    \caption{Data used to evaluate the thermal conductivity integral of PAI over temperature, produced by tracing plots from literature \cite{ThermalConductivityTorlon, Thermalexpansiontorlon} and stitching traced values together.}
    \label{fig:ConductivityIntegral}
\end{figure}

Figure \ref{fig:TorlonRegions} shows an engineering drawing of the PAI tube with regions labeled based on their temperature profile. To estimate the heat load on the CBGB cell due to thermal conduction, we assume the entire region of the Torlon that is in contact with the brass tube (Region C) is at cell temperature (2.40~K without helium flow, 2.51~K with 3.92~sccm helium flow), while the entire region in contact with the copper tube is at the heated fill line temperature (232~K without helium flow, 226~K with 3.92~sccm helium flow). For these calculations, we consider the steady-state condition without helium flow. By Fourier's Law of Heat Conduction,
\begin{equation}
    Q=-k(T)A \frac{dT}{dx}
    \label{eq:HeatConduction}
\end{equation}
where $k(T)$ is the temperature-dependent thermal conductivity of PAI. At steady state the heat flux $Q$ along the piece is constant. The conductive cross section $A$ for our piece is also constant, so we can rearrange and integrate along the piece.
\begin{align}
    \int_{0}^{L} \frac{Q}{A} dx &= -\int_{T_H}^{T_C} k(T) dT \\
    Q &= -\frac{A}{L}\int_{T_H}^{T_C} k(T) dT
    \label{eq:ConductionIntegration}
\end{align} 
We assume the thermal gradient exists only on Region I, whose length is $L=0.540$". Based on the drawing, cross sectional area is taken to be that of a cylindrical tube with an inner diameter of 0.2300" and a wall thickness of 0.010" (neglecting the taper from the drill bit). The thermal conductivity integral is evaluated from $T_H = 232$~K to $T_C=2.4$~K. To evaluate the integral, plots of the temperature-dependent thermal conductivity of PAI found in the literature \cite{ThermalConductivityTorlon, Thermalexpansiontorlon} were stitched together, traced, and numerically integrated using the trapezoidal approximation (with the traced data shown in Figure \ref{fig:ConductivityIntegral}. The resulting thermal conductivity integral is $29\, \mathrm{W} / \mathrm{m}\cdot\mathrm{K}$, resulting in an estimated heat load of $\sim$10~mW from heat conduction through the PAI tube.

Heat transfer also occurs via blackbody radiation from the portion of the injector inside the cell. The heat load $Q_{\mathrm{rad}}$ is given by the Stefan-Boltmann Law:
\begin{equation}
    Q_{\mathrm{rad}} = \epsilon \sigma A T_{\mathrm{injector}}^4
    \label{eq:StefBoltz}
\end{equation}
where we assume a worst case emissivity of $\epsilon=1$ for the Copper (C12200) surface of the heated fill line (in reality, the emissivity is highly dependent on the conditions of the surface, and for good mechanical and chemical polishing can be as low as $\epsilon\approx0.01$~\cite{databasemetallicmaterials}). The cylindrical surface of the tube (with an outer diameter of 0.0625" over a 1.9" length) and the annulus of the tip (with an outer diameter of 0.0625" and a 0.014" wall thickness) contribute to the area. We assume the entire fill line is at $T_{\mathrm{injector}}=232$~K. The resulting heat load is $Q_{\mathrm{rad}}\approx80$~mW.

\section{Heat Load Calibration} \label{sec:CellCalib}
\begin{figure}
    \centering
    \includegraphics[width=\columnwidth]{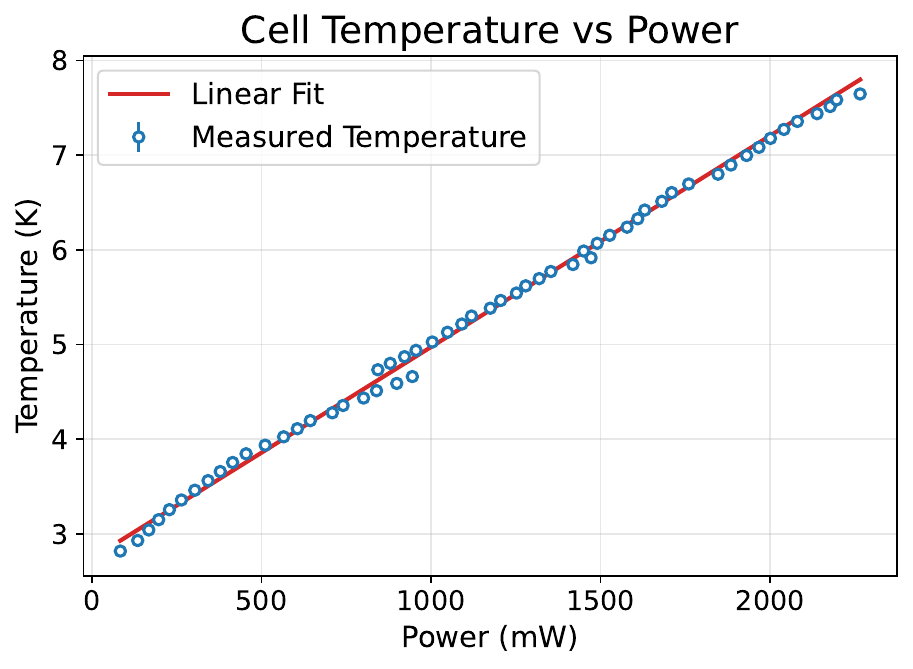}
    \caption{Relationship between cell temperature and applied heat load on cell.}
    \label{fig:CellCalib}
\end{figure}

The observed heat loads on the cell were experimentally determined based on the temperature of the cell. To find the relationship between applied heat load and cell temperature, a resistive heater\footnote{MP930 1k$\Omega$} was placed on the cell, diagonally opposite from the cell's thermometer\footnote{Lakeshore DT670 Silicon Diode}. The relationship between the heat load applied to the heater (reported by the power supply) and the cell temperature is shown in Figure \ref{fig:CellCalib}. The slope was found to be $2.23 \pm 0.02$~K/W, which corresponds to $448 \pm 4$~mW/K. For each applied heat load, we wait 5 minutes to allow the cell temperature to stabilize, then poll the temperature every 5 seconds for the next 5 minutes. The data in Figure \ref{fig:CellCalib} 
shows the averages and standard deviations of those temperatures for each applied power.

This relationship allows us to find the observed heat load for various operating conditions. Prior to installing the heated fill line, the base temperature of the cell was 2.18~K. For operation with \ce{SF6}, the heater on the VCR gland (shown in Figure \ref{fig:HFLinBeambox}) is set to 256~mW, resulting in an injector temperature of 232~K and a cell temperature of 2.4~K (without gas flow). Thus, the applied heat load on the cell is $\sim$98 mW. With 3.92~sccm of helium flow, the cell temperature rises to 2.51~K, corresponding to an additional heat load of $\sim$49~mW. When operating the heated fill line at 300~K, the cell temperature is 2.55~K, corresponding to a total heat load of  $\sim$166~mW. 

\section{Simulation and Analysis of $^{138}$\ce{BaF} Hyperfine Spectra}\label{sec:BaFsims}

The $X\rightarrow A$ transition in $^{138}$BaF has been well characterized by multiple high-resolution spectroscopy studies~\cite{PrecisionspectroscopyA2P-X2S,MolecularbeamopticalStark,HighResSpecBaF}. In particular, Ref.~\cite{PrecisionspectroscopyA2P-X2S} performed laser-induced fluorescence on cryogenic buffer gas beam with resolution near the natural linewidth of the $A{}^2\Pi_{1/2}$ state of $\Gamma/2\pi =3.46$~MHz. This study clearly resolved the hyperfine structure in both $A$ and $X$, and provided line assignments with <1~MHz precision. A comparison of their observed lines with ours from this study indicate agreement to $\lesssim\!5$~MHz, which agrees with our lock deviation when scanning. 

The transition frequencies can be modeled using an effective Hamiltonian approach~\cite{BrownCarrington} using parameters obtained from fits to experimental spectra. We use the effective Hamiltonian and parameters from Ref.~\cite{PrecisionspectroscopyA2P-X2S}, reproduced below:
\begin{align}
\hat{H}_{X} \;=&\;
\underbrace{B\,\mathbf{N}^{2}-D\,(\mathbf{N}^{2})^{2}}_{\text{Rotation}}
\;+\;\underbrace{\gamma\,\mathbf{N}\!\cdot\!\mathbf{S}}_{\text{Spin-Rotation}}
\\
&+\;\underbrace{b_F\,\mathbf{I}\!\cdot\!\mathbf{S}\;+\;\tfrac{c}{3}\,\sqrt{6}\,T^{2}_{0}(\mathbf{I},\mathbf{S})}_{\text{Magnetic Hyperfine (F)}}
\end{align}

\begin{align}
    \hat{H}_{A} \;=&\; \underbrace{T_v}_{\text{Origin}} \;+\;
\underbrace{A_{\mathrm{SO}}\,L_zS_z}_{\text{Spin–Orbit}}
\;+\;\underbrace{B\,\mathbf{N}^{2}-D\,(\mathbf{N}^{2})^{2}}_{\text{Rotation}}\\
&+\;\underbrace{
(p+2q)\!\sum_{\pm}\! e^{\mp 2i\phi}\,T^{2}_{\pm 2}(\mathbf{J},\mathbf{S})}_{\Lambda\text{-doubling}}\\
&-\; \underbrace{q\!\sum_{\pm}\! e^{\mp 2i\phi}\,T^{2}_{\pm 2}(\mathbf{J},\mathbf{J})}_{\Lambda\text{-doubling}}\\
&+\;\underbrace{h_{1/2}\,I_z L_z \;-\; d\,\sum_{\pm} \! e^{\mp 2i\phi} \, T^{2}_{\pm 2}(\mathbf{I},\mathbf{S})}_{\text{Magnetic Hyperfine (F)}}
\end{align}
Here, $\phi$ is the azimuthal angle in the $xy$ plane of the molecular frame, and the operator components are defined in the molecule frame, including the spherical tensors $T^k_q(...)$. Matrix elements were taken from Refs.~\cite{BrownCarrington,Hirota1985,MeasuringFundamentalSymmetry}. Effective Hamiltonian parameters are used from Ref.~\cite{PrecisionspectroscopyA2P-X2S}.

The cross-section for absorption on a given transition is proportional to the square of the transition dipole moment (TDM) between ground and excited states. We write the eigenstates after diagonalization as $|X,i\rangle$ and $|A,j\rangle$ respectively, with state indices $i$ and $j$. We perform a basis change~\cite{Hirota1985} to represent all states in the same basis (i.e. Hund's case). The TDM for the $X\rightarrow A$ electric dipole transition is given by
\begin{equation}
    D_{ij}=\langle X,i | T^1_{\pm1}(\mathbf{D})| A,j\rangle
\end{equation}
Here, $\mathbf{D} =-e\mathbf{r}$ is the electric dipole operator.  The TDM matrix elements can be evaluated by rotating from the molecule frame into the lab frame, and expressions can be found in Refs.~\cite{BrownCarrington,Hirota1985,MeasuringFundamentalSymmetry}. Since this is a ``perpendicular'' transition with $\Delta \Lambda=\pm 1$, it is driven by the $\pm1$ spherical tensor components of $\mathbf{D}$ in the molecule frame. The resulting normalized intensities for the three lines were calcualted to be approximately  (22\%, 44\%, 22\%) for the $F^{\prime\prime}=(1,1,0)\rightarrow F^\prime=(0,1,1)$ transitions. We note that spin-orbit and Coriolis mixing can cause interference and anomlaous transition intensities~\cite{PrecisionspectroscopyA2P-X2S,CaltechRaX,Jadbabaie2023}. Indeed, the front of cell fits in Figure~\ref{fig:spectrum} seem to indicate the $F^{\prime\prime} =0\rightarrow F^\prime=1$ transition is weaker than the ratios used in the simulation.

We construct simulated spectra by plotting Voigt lineshapes for each initial and final state with nonzero $|D_{ij}|^2$. The Lorentzian width is fixed at the natural linewidth $\gamma/2\pi=3.46$~MHz, and a single Gaussian width is used for all transitions, while the Voigt amplitude is scaled by $|D_{ij}|^2$ for each transition. Since we are working with an isotropic, unpolarized sample, we sum over all $M$-sublevels in the ground $N''=0^+$ and excited $J'=1/2^-$ state manifolds. The result is three lines: $F^{\prime\prime}=(1,1,0)\rightarrow F^\prime=(0,1,1)$.

For the fits to the data, we use a three-line Voigt fit. We fix the line splittings to the simulated values, and the ratio of the line heights to the predictions of the $|D_{ij}|^2$ calculation. The remaining free parameters are then an overall frequency offset, the overall height, the overall background, and the Gaussian width. Our overall offset fit value is within our quoted uncertainty of 5 MHz when scanning. 

\end{document}